{}%disable hyper at arxiv
{}%use pdflatex at arxiv

\documentclass[prd,reprint,superscriptaddress,preprintnumbers]{revtex4-1}

\newif\ifpublic\publictrue
\usepackage{fancyhdr}
\ifpublic\else
\fi
\newif\ifworking\workingtrue

\usepackage{mathtools,mathrsfs,amsbsy,amssymb,latexsym,amsfonts,amscd,
	amsmath,, multirow}
\usepackage{bm}

\usepackage{graphicx,caption,subcaption}
\usepackage[usenames,dvipsnames]{color}
\usepackage[normalem]{ulem}
\usepackage{natbib}
\usepackage{xcolor}
\usepackage{comment}
\definecolor{linkcolor}{rgb}{0,0,0.6}
\usepackage[ pdftex,colorlinks=true,
pdfstartview=FitV,
linkcolor= linkcolor,
citecolor= linkcolor,
urlcolor= linkcolor,
hyperindex=true,
hyperfigures=false]
{hyperref}
\newcommand{\uA}{{\underline{A}}}
\newcommand{\uB}{{\underline{B}}}

\newcommand{\uD}{{\underline{D}}}

\newcommand{\uM}{{\underline{M}}}
\newcommand{\uN}{{\underline{N}}}
\newcommand{\uP}{{\underline{P}}}
\newcommand{\cI}{{\cal I}}
\newcommand{\cJ}{{\cal J}}
\newcommand{\cK}{{\cal K}}
\newcommand{\cL}{{\cal L}}

\newcommand{\cT}{{\cal T}}
\newcommand{\n}{\ell}

\def\showkeysrefformat#1{{\normalfont\tiny\ttfamily#1}}
\makeatletter
\def\SK@@ref#1>#2\SK@{%
	{\@inlabelfalse\leavevmode\vbox to\z@{%
			\vss\SK@refcolor\rlap{\vrule\raise .75em%
				\hbox{\showkeysrefformat{#2}}}}}}
\makeatother

\allowdisplaybreaks[3]

\begin{document}
\title{Stable non-supersymmetric Anti-de Sitter vacua of massive IIA supergravity}

\author{Adolfo Guarino}
\email{adolfo.guarino@uniovi.es}
\affiliation{Departamento de F\'isica, Universidad de Oviedo, Avda. Federico Garc\'ia Lorca 18, 33007 Oviedo, Spain}
\affiliation{Instituto Universitario de Ciencias y Tecnolog\'ias Espaciales de Asturias (ICTEA), Calle de la Independencia 13, 33004 Oviedo, Spain}

\author{Emanuel Malek}
\email{emanuel.malek@physik.hu-berlin.de}
\affiliation{Institut f\"ur Physik, Humboldt-Universit\"at zu Berlin, IRIS Geb\"aude, Zum Gro{\ss}en Windkanal 6, 12489 Berlin, Germany}

\author{Henning Samtleben}
\email{henning.samtleben@ens-lyon.fr}
\affiliation{Univ Lyon, Ens de Lyon, Univ Claude Bernard, CNRS,	Laboratoire de Physique, F-69342 Lyon, France}

\preprint{HU-EP-20/34}

\begin{abstract}
Stable non-supersymmetric Anti-de Sitter (AdS) vacua of string theory are widely believed not to exist. In this letter, we analytically compute the full bosonic Kaluza-Klein spectrum around the ${\rm G_2}$-invariant non-supersymmetric $\textrm{AdS}_4$ solution of massive IIA supergravity, and show that it is perturbatively stable. We also provide evidence that six other non-supersymmetric $\textrm{AdS}_4$ solutions of massive IIA supergravity are perturbatively stable. Since previous studies have indicated that these AdS vacua may also be non-perturbatively stable, our findings pose a challenge to the Swampland Conjecture.	
\end{abstract}

\maketitle

Constructing stable non-supersymmetric vacua of string theory is one of its foremost challenges, and crucial for applications to phenomenology, cosmology and holography. Unlike their supersymmetric counterparts, such vacua are not protected by supersymmetry arguments and positive mass theorems \cite{Gibbons:1983aq}. The stability of non-supersymmetric Anti-de Sitter (AdS) vacua is particularly interesting, since they would provide a crucial tool for applying holography to realistic systems in condensed matter or indeed QCD. Moreover, non-supersymmetric AdS vacua might be a stepping stone to understanding time-dependent de Sitter solutions in string theory.

The stability of non-supersymmetric AdS vacua has proven particularly contentious in recent years. Various arguments have been put forward based on the Weak Gravity Conjecture \cite{ArkaniHamed:2006dz} that all non-supersymmetric AdS vacua of string theory are unstable, ultimately leading to the AdS Swampland Conjecture \cite{Ooguri:2016pdq}. Moreover, despite years of trying, no fully-fledged case of a stable non-supersymmetric AdS vacuum is known in string theory. Indeed, a powerful method for constructing non-supersymmetric AdS vacua involves uplifting non-supersymmetric solutions of lower-dimensional supergravities via a consistent truncation. However, all but a small number of these AdS solutions are already unstable within the lower-dimensional supergravity, with some of the scalar fields violating the Breitenlohner-Freedman (BF) bound \cite{Breitenlohner:1982jf}. Notable exceptions to this are the non-supersymmetric $\textrm{SO}(3) \times \textrm{SO}(3)${{-invariant}} AdS$_4$ vacuum of 4-dimensional ${\cal N}=8$ SO$(8)$ gauged supergravity \cite{Warner:1983vz}, obtained by consistent truncation of 11-dimensional supergravity on $S^7$ \cite{deWit:1982ig}, and seven non-supersymmetric AdS$_4$ vacua of 4-dimensional ${\cal N}=8$ ISO$(7)$ gauged supergravity (see appendix A of \cite{Guarino:2020jwv} for a summary and original references), obtained by consistent truncation of massive IIA supergravity on $S^6$ \cite{Guarino:2015jca}. Remarkably, for these non-supersymmetric AdS$_4$ vacua all 70 scalar fields in the $\mathcal{N}=8$ supergravity multiplet have masses above the BF bound \cite{Fischbacher:2010ec,Guarino:2020jwv}.

Although these vacua are stable in four dimensions, their higher-dimensional stability is far from guaranteed. For example, the Kaluza-Klein (KK) spectrum around such vacua could contain tachyonic scalars violating the BF bound, or the AdS vacua may exhibit non-perturbative instabilities. Understanding the higher-dimensional stability has long been a challenge for these vacua, since both the computation of Kaluza-Klein spectra and the systematic search for non-perturbative instabilities, see e.g.\ \cite{Apruzzi:2019ecr}, used to be notoriously difficult. However, within the last year, a new and powerful method based on Exceptional Field Theory (ExFT) \cite{Hohm:2013pua} was developed in \cite{Malek:2019eaz,Malek:2020yue} for computing Kaluza-Klein spectra of vacua of maximal gauged supergravities. At the same time, \cite{Bena:2020xxb} showed that probe branes can be used to easily search for signals, dubbed brane-jet instabilities, of non-perturbative instabilities \cite{Apruzzi:2019ecr}. These new methods were used to show that the $\textrm{SO}(3) \times \textrm{SO}(3)$-invariant AdS$_4$ vacuum has tachyonic Kaluza-Klein modes \cite{Malek:2020mlk} and suffers from M$2$-brane-jet instabilities \cite{Bena:2020xxb}, with brane-jet instabilities also arising for other AdS vacua \cite{Suh:2020rma}.

In this letter, we will investigate the Kaluza-Klein spectrum of the seven non-supersymmetric AdS$_4$ vacua of the maximal ISO$(7)$ supergravity, which correspond to compactifications of massive IIA supergravity on various deformed $S^6$ spheres, and are stable within the four-dimensional supergravity. In particular, we focus on the vacuum with the largest bosonic symmetry, $\textrm{G}_2$. Remarkably, and in contrast to the $\textrm{SO}(3) \times \textrm{SO}(3)$ symmetric AdS$_4$ vacuum of the SO$(8)$ theory, the AdS$_4$ vacuum with $\textrm{G}_{2}$ symmetry has shown to be D$p$-brane-jet stable for $p=2,4,6,8$, while some other non-perturbative channels have also been ruled out \cite{Guarino:2020jwv}. Moreover, a partial search for the other six AdS$_4$ vacua has also found no signs of D$2$-brane-jet instabilities \cite{Guarino:2020jwv}. Finally, because the compactification spaces have $S^6$ topology, supported by fluxes, these vacua do not suffer from decays due to ``bubbles of nothing'' of the type argued in \cite{Ooguri:2017njy}. This suggests that these vacua may be non-perturbatively stable, making it even more important to study their perturbative Kaluza-Klein stability in light of the Weak Gravity and Swampland conjectures. This is what we set up to do in this letter. In particular, we compute the full Kaluza-Klein spectrum of the $\textrm{G}_2$-invariant vacuum, providing the first analytic spectrum of a non-supersymmetric vacuum, and use this to prove its perturbative stability. Moreover, we collect numerical evidence for the perturbative stability of the six other AdS$_{4}$ vacua.

In \cite{Malek:2019eaz,Malek:2020yue}, ExFT was used to derive mass matrices for the Kaluza-Klein spectrum around any vacuum of ${\cal N}=8$ gauged supergravities in four and five dimensions that arises from a consistent truncation of 10-/11-dimensional supergravity. Let us briefly review some of the salient features of ExFT and Kaluza-Klein spectroscopy.

ExFT is a reformulation of 10-/11-dimensional supergravity, which unifies the metric and flux degrees of freedom
within a manifestly E$_{7(7)}$ covariant formulation. Its bosonic sector
\begin{equation}
	\begin{split}
	\left\{g_{\mu\nu}, {\cal M}_{MN},{\cal A}_\mu{}^M \right\}	\,, \quad&
	\mu=0, \dots,3\,,\\
	&{}
	M=1, \dots, 56
	\,,
	\label{fields}
	\end{split}
\end{equation}
consists of an external and an internal metric $g_{\mu\nu}$, ${\cal M}_{MN}$, respectively, with the latter parameterising the coset space E$_{7(7)}/{\rm SU}(8)$, together with vector fields, ${\cal A}_\mu{}^M$, transforming in the ${\bf 56}$ of the group E$_{7(7)}$.

As shown in \cite{Malek:2019eaz,Malek:2020yue}, a general Kaluza-Klein fluctuation around a vacuum that uplifts from four-dimensional gauged supergravity can be expressed as a product of the modes of the consistent truncation, with a complete basis of functions on the compactification manifold. A powerful feature of this method is that this complete basis of functions can be chosen to be the scalar harmonics, ${\cal Y}^\Sigma$, of the compactification with a metric that preserves the largest possible symmetry group, ${\rm G_{max}}$, in the lower-dimensional gauged supergravity. In the case of the maximal ISO(7) supergravity investigated in this letter, the internal space topology is $S^{6}$, the largest possible symmetry group is $\textrm{G}_{\textrm{max}}=\textrm{SO(7)}$, and we can choose the ${\cal Y}^\Sigma$ to be the scalar harmonics on the round $S^6$.

The fluctuation Ansatz of the ExFT fields \eqref{fields} around an AdS$_{4}$ vacuum is given by \cite{Malek:2019eaz,Malek:2020yue}
\begin{equation} \label{eq:FluctAnsatz}
	\begin{split}
		g_{\mu\nu}(x,y) & = \rho^{-2} \, \Big( \mathring{g}_{\mu\nu}(x) + \displaystyle\sum_{\Sigma} \mathcal{Y}^{\Sigma} \, h_{\mu\nu , \Sigma}(x) \Big) \,, \\
		\mathcal{A}_{\mu}{}^{M}(x,y) & =  \rho^{-1} \, (U^{-1})_{\uA}{}^{M} \, \displaystyle\sum_{\Sigma} \mathcal{Y}^{\Sigma} \, A_{\mu}{}^{\uA , \Sigma}(x) \,, \\
		\mathcal{M}_{MN}(x,y) & = U_{M}{}^{\uA} 
		U_{N}{}^{\uB} \, \Big( \delta_{\uA\uB} + {\cal P}_{\uA\uB,I}\displaystyle\sum_{\Sigma} \mathcal{Y}^{\Sigma}  j_{I , \Sigma}(x) \Big) \,,
	\end{split}
\end{equation}
where the Kaluza-Klein fluctuations for the metric, vector fields and scalars are labeled by $h_{\mu\nu,\Sigma}(x)$, $A_{\mu}{}^{\uA,\Sigma}$, and $j_{I,\Sigma} \in \mathfrak{e}_{7(7)} \ominus \mathfrak{su}(8)$, respectively. The latter appear under projection ${\cal P}_{\uA\uB,I}$, with $I=1,\ldots,70$, resulting from the expansion of the group element $\mathcal{M}_{MN}$. On the other hand, $\rho(y) \in \mathbb{R}^{+}$ and $U_{M}{}^{\underline{M}}(y)\in \textrm{E}_{7(7)}$ denote the scaling function and the twist matrix, respectively, encoding the consistent truncation to ${\cal N}=8$ gauged supergravity \cite{Hohm:2014qga}. In the fluctuation Ansatz \eqref{eq:FluctAnsatz}, the twist matrix appears dressed with the scalar matrix of the four-dimensional supergravity, ${\cal V}_{\underline{M}}{}^{\underline{A}} \in \textrm{E}_{7(7)}/\mathrm{SU}(8)$, evaluated at the scalar configuration specifying the vacuum of the maximal $D=4$ supergravity, i.e.
\begin{equation}
\label{dressed_U}
	U_{M}{}^{\uA}(y)=U_{M}{}^{\underline{M}}(y) \, \mathcal{V}_{\underline{M}}{}^{\underline{A}} \,.
\end{equation}

The fluctuation Ansatz \eqref{eq:FluctAnsatz} induces mass matrices for the Kaluza-Klein spectrum which are entirely expressed through the embedding tensor of the ${\cal N}=8$ gauged supergravity, $X_{\underline{MN}}{}^{\underline{P}}$, and the linear action, $\cT_{\underline{M}}{}^\Sigma{}_\Omega$, of the $\textrm{G}_{\textrm{max}}$ Killing vector fields on the scalar harmonics, both dressed by the scalar matrix $\mathcal{V}_{\underline{M}}{}^{\underline{A}}$.
The matrices $\cT_{\underline{M}}{}^\Sigma{}_\Omega$ are explicitly defined as
\begin{equation}
	L_{{\cal K}_{\underline{M}}} {\cal Y}^\Sigma = - \cT_{\underline{M}}{}^{\Sigma}{}_\Omega\, {\cal Y}^\Omega \,,
\end{equation}
where ${\cal K}_{\underline{M}}$ are the Killing vectors, generating ${\rm G_{max}}$, which can be extracted from the twist matrix $U_{M}{}^{\underline{M}}$\,. In turn, the dressed objects are defined as
\begin{equation}
	\label{T-tensor}
	\begin{split}
	X_{\underline{AB}}{}^{\underline{C}} &= (\mathcal{V}^{-1})_{\uA}{}^{\underline{M}} \, (\mathcal{V}^{-1})_{\uB}{}^{\underline{N}}  \, X_{\underline{MN}}{}^{\underline{P}} \,  \mathcal{V}_{\underline{P}}{}^{\underline{C}} \,, \\
	{\cal T}_{\uA}{}^{\Sigma}{}_\Omega &= (\mathcal{V}^{-1})_{\uA}{}^{\underline{M}} \, \cT_{\uM}{}^{\Sigma}{}_{\Omega} \,.
	\end{split}
\end{equation}

The mass matrices are obtained by linearizing the ExFT field equations with the fluctuation ansatz \eqref{eq:FluctAnsatz} \cite{Malek:2019eaz,Malek:2020yue}. For the purpose of this letter, we give the scalar mass matrix in a yet more compact form:
\begin{equation}
\begin{split}
{\cal M}^{{\rm (scalar)}}_{I\Sigma,J\Omega}  
&=
{\cal M}^{(0)}_{IJ}\,\delta_{\Sigma\Omega}
+{\delta}_{IJ}\,\mathbb{M}^{(2)}_{\Sigma\Omega} 
+{\cal N}_{IJ}{}^{\underline{C}}\,{\cal T}_{\underline{C},\Sigma\Omega}
\\&\quad
-\tfrac16(\Pi^T \Pi)_{I\Sigma,J\Omega} 
\,.
\end{split}
\label{scalar_mass}
\end{equation}
Here, ${\cal M}^{(0)}_{IJ}$ and $\mathbb{M}^{(2)}_{\Sigma\Omega} = -({\cal T}_{\underline{A}}{\cal T}_{\underline{A}})_{\Sigma\Omega}$ are the mass matrices of the four-dimensional supergravity scalars and of the spin-2 fluctuations, respectively. The matrices in the last two terms are given by
\begin{equation}
\begin{split}
{\Pi}_{\underline{A}\Sigma,I\Omega}
&=
\delta_{\Sigma\Omega}\,
X_{\underline{AC}}{}^{\underline{D}} \, {\cal P}_{\underline{CD},I} 
- 12\,{\cal P}_{\underline{AD},I}\,  {\cal T}_{\underline{D}\,\Omega}{}^{\Sigma} 
\,,
\\
{\cal N}_{IJ}{}^{\underline{C}}
&=
-4\left(X_{\underline{CA}}{}^{\underline{B}}+12\,X_{\underline{AB}}{}^{\underline{C}}\right)
{\cal P}_{\uA\uD}{}^{[I} 
{\cal P}_{\uB\uD}{}^{J]}
\,.
\end{split}
\end{equation}
In particular, the matrix ${\Pi}_{\underline{A}\Sigma,I\Omega}$ features in the linearized covariant scalar derivatives, $\partial_\mu j_{I,\Sigma} -{\cal A}_{\mu}{}^{\underline{C}\Omega} \,{\Pi}_{\underline{C}\Omega,I\Sigma}$ and thus encodes the projection onto the Goldstone scalars. Accordingly, it is orthogonal to the mass matrix \eqref{scalar_mass} and encodes the vector mass matrix as ${\cal M}^{{\rm (vector)}}=\Pi \,\Pi^T$\,.

Let us now specialize to the specific ISO$(7)$ gauged supergravity whose vacua we are interested in. In terms of SL$(7) \subset \textrm{E}_{7(7)}$, defined by the branching of the $\mathbf{56}$ as
\begin{equation}
	A^{\uM} \rightarrow 
	\left\{ A^{[ab]},\, A^{a8},\, A_{[ab]},\, A_{a8} \right\} \,,
\end{equation}
with $a=1,\ldots,7$, the only non-zero components of the embedding tensor are
\begin{equation}
\label{X_ISO(7)}
	X_{\uM\uN}{}^{\uP} = \left\{ \begin{array}{l}
		X_{ab \, cd}{}^{ef} = - X_{ab\,}{}^{ef}{}_{cd} = -8 \, \delta_{[a}^{[e} \delta^{\phantom{[a}}_{b][c} \delta_{d]}^{f]} \,,  \\[2mm]
		X_{ab \, c8}{}^{d8} = - X_{ab\,}{}^{d8}{}_{c8} =  -2 \, \delta_{[a}^{d} \delta^{\phantom{[a}}_{b]c}  \,, \\[2mm]
		X_{c8 \, ab}{}^{d8} = - X_{c8\,}{}^{d8}{}_{ab} =  -2 \, \delta^{\phantom{[a}}_{c[a} \delta_{b]}^{d}  \,,  \\[2mm]
		X^{c8\, }{}_{ab}{}^{d8} = - X^{c8 \, d8}{}_{ab} =  2 \, c \, \delta^{c}_{[a} \delta^{d}_{b]} \ .
	\end{array} \right.
\end{equation}
Here, $c$ is a magnetic parameter that renders the gauging of dyonic type \cite{deWit:2005ub,Dall'Agata:2014ita} and which is identified with the Romans mass parameter \cite{Romans:1985tz} in ten dimensions.

For the fluctuation Ansatz, we will be using the scalar harmonics on the round $S^6$, which can be expressed as symmetric traceless polynomials in the elementary harmonics ${\cal Y}^a$ satisfying ${\cal Y}^a\, {\cal Y}^b\, \delta_{ab} = 1$. These are given by
\begin{equation}
	{\cal Y}^\Sigma = \left\{ 1,\, {\cal Y}^a,\, {\cal Y}^{a_1a_2},\, \ldots,\, {\cal Y}^{a_1\ldots a_\n},\, \ldots \right\} \,,
\label{harmonics}
\end{equation}
with ${\cal Y}^{a_1 \ldots a_\n} \equiv {\cal Y}^{(\!(a_1} \ldots {\cal Y}^{a_\n)\!)}$ and $(\!(\, )\!)$ denotes traceless symmetrisation. Therefore, the $\cT_{\uM}{}^{\Sigma}{}_{\Omega}$ now simply become the SO$(7)$ generators in the $\left[\n,0,0\right]$ representation, with the only non-zero components given by
\begin{equation}
\label{T_SO(7)_generators}
	\cT_{ab}{}^{c_1\ldots c_\n}{}_{d_1\ldots d_\n} = 4\, \n\, \delta_{[a}{}^{(\!(c_1} \delta_{b](\!(d_1} \delta_{d_2}{}^{c_2} \ldots 
	\delta_{d_\n)\!)}{}^{c_\n)\!)} \,.
\end{equation}
We stress once more that the harmonics \eqref{harmonics} constitute an appropriate basis for the analysis of all vacua of this theory, regardless of the specific $S^6$ metric occurring at the vacua. With the data in \eqref{X_ISO(7)} and \eqref{T_SO(7)_generators} specific to the maximal ISO(7) supergravity, we can now proceed and compute the KK spectrum of the metric, vector and scalar fluctuations around a given AdS$_{4}$ vacuum specified by the scalar matrix ${\cal V}_{\underline{M}}{}^{\underline{A}}$ via the dressed tensors in \eqref{T-tensor}.

The maximal ISO(7) supergravity possesses a rich structure of AdS$_4$ vacua. To date, $60$ such vacua have been identified, up to discrete degeneracies, (see appendix~A of \cite{Guarino:2020jwv}), all of which are contained within a $\mathbb{Z}_{2}^3$-invariant sector of the theory \cite{Guarino:2019snw}. This sector describes an $\mathcal{N}=1$ supergravity coupled to seven chiral multiplets with complex scalar components $z_{i}$ ($i=1,\ldots,7$). Thus, the $56$-bein entering the fluctuation Ansatz \eqref{eq:FluctAnsatz} via the dressed twist matrix in \eqref{dressed_U} parameterizes a coset subspace
\begin{equation}
\label{56-bein}
\mathcal{V}_{\underline{M}}{}^{\underline{A}}(z_{i})  \in  \left[\textrm{SL}(2)/\textrm{SO}(2)\right]^7 \subset \textrm{E}_{7(7)}/\textrm{SU}(8) \ .
\end{equation}
The K\"ahler potential $K(z_{i},\bar{z}_{i})$ and holomorphic superpotential $W(z_{i})$ for the complex $z_{i}$ are given by
\begin{equation}
	\label{N=1_K&W}
	\begin{split}
		K(z_{i},\bar{z}_{i}) &=  - \displaystyle\sum_{i=1}^7   \log[-i(z_{i}-\bar{z}_{i})] \,, \\
		W(z_{i}) &= 2  g \, \left[ \, z_{1} z_{2} z_{3} +  z_{1} z_{6} z_{7} +  z_{2} z_{5} z_{7} + z_{3} z_{5} z_{6}\right. \\
		& \quad \left. \qquad + (  z_{1} z_{5} +  z_{2} z_{6} + z_{3} z_{7} ) \,  z_{4} \, + c  \, \right]  \ ,
	\end{split}
\end{equation}
where $g$ is the coupling constant in the four-dimensional supergravity that relates to the (inverse) radius of $S^6$.
\begingroup
\squeezetable
\begin{table*}[t]
	\renewcommand{\arraystretch}{1.5}
	\begin{tabular}{cccc}
		\noalign{\hrule height 1.5pt}
		ID &		$\textrm{G}_{\rm res}$ \,&\, $c^{-\frac{1}{3}}  \, z_{i}$  & $g^2 \, c^{-\frac{1}{3}}  \, L^2$  \\
		\noalign{\hrule height 1.5pt}
		(a) &			$\textrm{SU}(3)$ \,&\,  $z_{1,2,3} = -0.2698+0.7329 \,  i \,, \quad
		z_{4,5,6,7} = 0.4915 + 0.6618 \, i$ & $0.512522$ \\
		(b) &			$\textrm{SU}(3)$ \,&\,  $z_{1,2,3} = -0.4547 + 0.8376 \,  i  \,, \quad
		z_{4,5,6,7} = 0.3348 + 0.6012 \, i $  & $0.511579$ \\
		(c) &			$\textrm{SO}(4)$ \,&\,  $z_{1,2,3,5,6,7} =  - 0.4123 + 0.6512\,  i  \,, \quad
		z_{4} = - 0.0678 + 1.1466 \, i $   & $0.510363$ \\
		(d) &			$\textrm{U}(1) \times \textrm{SO}(3)$ \,&\,  $z_{1,7} = -0.5061 + 0.7328 \,  i \,, \quad
		z_{2,3,5,6} = -0.3520 + 0.6240 \, i  \,,\quad
		z_{4} = -0.2732 + 0.9609 \, i $   & $0.511595$ \\
		(e) &			$\textrm{U}(1) \times \textrm{SO}(3)$ \,&\,  $z_{1,7} = -0.4869 + 0.7937 \,  i  \,,\quad
		z_{2,3,5,6} = -0.3382 + 0.6059 \, i \,,\quad
		z_{4} = -0.3736 + 0.9082 \, i $   & $0.511536$ \\
		(f) &			 $\textrm{SO}(3)$ \,&\,  $z_{1,7} = -0.5065 + 0.7285 \,  i \,, \quad
		z_{2,6} = -0.3663 + 0.6256 \, i  \,, \quad
		z_{3,5} = -0.3403 + 0.6259  \, i \,, \quad z_{4} = -0.2671 + 0.9623 \, i$  &  $0.511594$ \\
		\noalign{\hrule height 1.5pt}
	\end{tabular}
	\caption{Identifier, residual symmetry group, location and AdS radius of the six non-supersymmetric and BF-stable AdS$_4$ vacua of the maximal ISO(7) supergravity with $\textrm{SU}(3)$, $\textrm{SO}(4)$, $\textrm{U}(1) \times \textrm{SO}(3)$ and $\textrm{SO}(3)$ symmetry.}
	\label{Table:six_vacua}
\end{table*}
\endgroup

This truncated sector contains 7 non-supersymmetric AdS$_4$ vacua, yet stable within $D=4$ supergravity. I.e.\ all 70 scalars from the 
${\cal N}=8$ supergravity multiplet have masses above the BF bound.  At these vacua, the ISO(7) symmetry of the maximal theory is broken to $\rm G_{res}$ \cite{Borghese:2012qm,Guarino:2015qaa,Guarino:2019jef} and the mass spectra organize into representations of $\rm G_{res}$.

Employing the techniques reviewed above, we now compute the full bosonic Kaluza-Klein spectra around these non-supersymmetric AdS$_{4}$ vacua. Let us first focus on the $\textrm{G}_{2}$-invariant vacuum located at
\begin{equation}
\label{AdS4_G2}
z_{1,2,3,4,5,6,7} = 2^{-\frac{4}{3}} \, (-1 + \sqrt{3} \, i) \, c^{\frac{1}{3}}\ ,
\end{equation}
with AdS radius $L^2=2^{-\frac{10}{3}} \, 3^{\frac{3}{2}} \, g^{-2} \, c^{\frac{1}{3}}$, and which was first constructed directly in 10 dimensions in \cite{Lust:2008zd}. The mass spectrum of bosonic fields in the maximal supergravity was presented in \cite{Borghese:2012qm} and a first rudimentary study of its higher-dimensional stability was performed in \cite{Cassani:2009ck}. Here we will compute its full bosonic Kaluza-Klein spectrum by exploiting the large residual symmetry group ${\rm G}_2$ to analytically evaluate the mass matrices \eqref{scalar_mass}.

The normalized spin-2 mass matrix reduces to the $\textrm{G}_2$ Casimir operator
\begin{equation}
L^2\,\mathbb{M}^{(2)}_{\Sigma\Omega}
=
-L^2\,({\cal T}_{\underline{A}}{\cal T}_{\underline{A}})_{\Sigma\Omega}
=\frac32 \, \left( {\rm Cas}_{G_2} \right)_{\Sigma\Omega}
\,,
\end{equation}
whose eigenvalues are given by \cite{Okubo:1977mx}
\begin{equation}
	\label{Casimir_G2}
	C_{[n_1,n_2]} = \frac13\left(5\,n_1+n_1^2+9\,n_2+3\,n_1n_2+3\,n_2^2 \right) \,,
\end{equation}
on the $\left[n_1,n_2\right]$ representation. The spin-2 fluctuations at level $\n$ transform in the $\left[\n,0\right]$ representation, which gives the mass eigenvalues $\frac12 \, \ell \,(\ell+5)$, in accordance with the results in \cite{Pang:2017omp}.

For the scalar Kaluza-Klein modes, the first three terms of the (normalized) mass matrix \eqref{scalar_mass} can be shown to combine into
\begin{equation}
6\,\delta_{IJ}\,\delta_{\Sigma\Omega}
+3\,({\rm Cas}_{G_2})_{\Sigma\Omega}\,\delta_{IJ}
-\frac32\,({\rm Cas}_{G_2})_{I\Sigma,J\Omega}
\,,
\end{equation}
where the two Casimir operators act on the representations of the spin-2 harmonics and on the scalar fluctuations $j_{I,\Sigma}$, respectively. In contrast, the last term in \eqref{scalar_mass} only affects the Goldstone scalars and ensures that these appear with zero eigenvalues. Putting this together with \eqref{Casimir_G2}, we find that the normalized scalar masses at level~$\ell$ allow for a surprisingly compact expression in terms of the quadratic $\textrm{G}_2$ Casimir eigenvalues as
\begin{equation}
\label{Mass_G2_spin0_spin1}
	M^2_{[n_1,n_2]_{\ell}}  L^{2} = (\ell+2)\,(\ell+3)-\frac32\,C_{[n_1,n_2]} \,,
\end{equation}
where $[n_1,n_2]_{\ell}$ denotes the  $\textrm{G}_2$ representation that the Kaluza-Klein modes at level $\n$ appear in. The form of the scalar mass matrix \eqref{scalar_mass} implies that this mass formula also extends to the spin-1 sector. In particular, when evaluated for the first two levels, it reproduces the vector masses computed in \cite{Varela:2020wty}.

For the spin-$0$ Kaluza-Klein modes, the relevant representations at level $\n$ are the product of those of level $0$, $2 \cdot \left( [2,0] \oplus [1,0] \oplus [0,0] \right)$, with $[\n,0]$, taking care to remove the Goldstone bosons. For generic $\n$, this gives
\begin{equation}
\label{G2_spin0_irreps}
	\begin{split}
		& 2\cdot [ \ell-4,2] \oplus [ \ell-3,1] \oplus [ \ell-3,2] \oplus 2\cdot [ \ell-2,0] \\
		\oplus &\, 3\cdot [ \ell-2,1] \oplus 2\cdot [ \ell-2,2]  \oplus [ \ell-1,0] \oplus 3\cdot [ \ell-1,1] \\
		\oplus &\, 5\cdot [\ell,0] \oplus  [\ell,1] \oplus [\ell+1,0] \oplus 2\cdot [\ell+2,0] \,,
	\end{split}
\end{equation}
with care needed for the degeneracies at low $\n$. Similarly, the spin-$1$ Kaluza-Klein modes at generic level $\ell$ transform in the representations
\begin{equation}
\label{G2_spin1_irreps}
	\begin{split}
	&\; [ \ell-3,1] \oplus [ \ell-3,2] \oplus 3\cdot[ \ell-2,1] \oplus 3\cdot[ \ell-1,0] \\
	 \oplus & \, 3\cdot[ \ell-1,1] \oplus 2\cdot[\ell,0] \oplus [\ell,1] \oplus 3\cdot[\ell+1,0] \,,
	\end{split}
\end{equation}
again with care needed for the degeneracies at low $\n$.
Evaluation of the mass formula \eqref{Mass_G2_spin0_spin1} on the list of $\textrm{G}_{2}$ representations in \eqref{G2_spin0_irreps} for the tower of spin-$0$ Kaluza-Klein modes, shows that the masses increase with increasing Kaluza-Klein level $\ell$ and are positive. This proves the perturbative stability of the non-supersymmetric $\textrm{G}_{2}$ vacuum.

\begin{figure*}[t]
\begin{center}
\includegraphics[width=\textwidth]{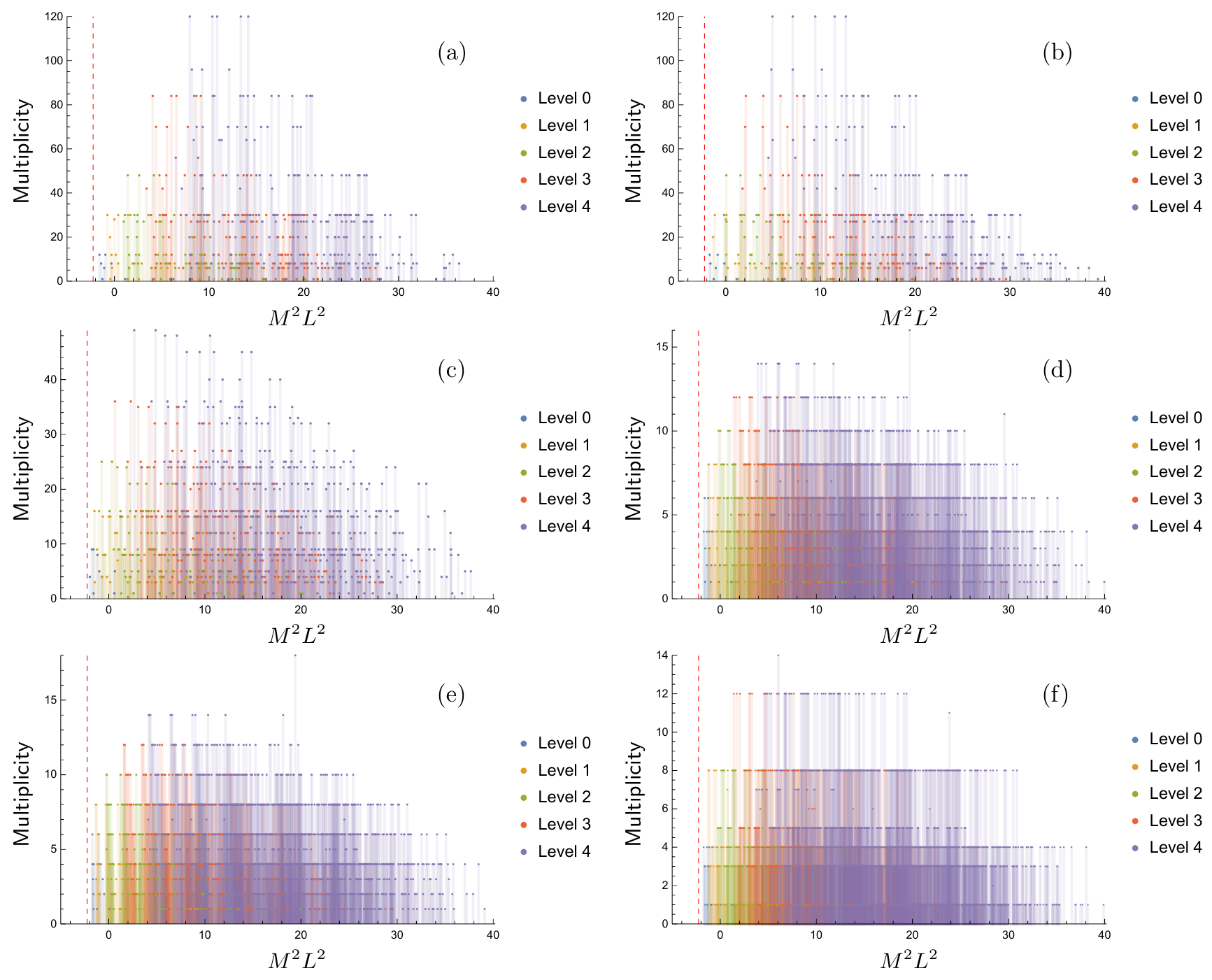}
\end{center}
\caption{Normalized masses $M^2 L^2$ and multiplicities for the first four KK levels of spin-$0$ fluctuations around the six non-supersymmetric and BF stable $\textrm{AdS}_{4} \times {S}^6$ solutions of massive IIA supergravity given in Table~\ref{Table:six_vacua}. The subfigure labels (a), (b), (c), (d), (e), (f) correspond to the ID's of the different vacua in Table~\ref{Table:six_vacua}. The dashed red line marks the BF bound $M^2 L^2 = - \frac94$.}
\label{Fig:KK_modes}
\end{figure*}

We finally turn to the other six non-supersymmetric $\textrm{AdS}_{4}$ solutions of massive IIA supergravity with smaller $\rm G_{res}$, which correspond to compactifications of massive IIA supergravity on various deformed $S^6$ spheres and are summarized in Table~\ref{Table:six_vacua}.  The small residual symmetry groups $\textrm{G}_{\rm res}$ do not allow us (for the moment) to analytically resolve these mass spectra. Instead, we revert to a numerical evaluation of the mass matrices \eqref{scalar_mass} up to and including KK level $\ell=4$, with results displayed in Figure~\ref{Fig:KK_modes}. Remarkably, the analysis shows that all scalars on these levels are perturbatively stable with their normalized masses lying above the BF bound. Moreover, as is pictured in Figure~\ref{Fig:KK_modes}, the lowest-lying normalized masses at each Kaluza-Klein level $\ell$ increase monotonically with the level, thus suggesting that the full spectra will also be stable.

To summarize, we have analytically computed the full Kaluza-Klein spectrum of the non-supersymmetric and $\textrm{G}_2$-invariant $\textrm{AdS}_4 \times S^{6}$ background of massive IIA supergravity, resulting in the closed mass formula \eqref{Mass_G2_spin0_spin1} encoding the entire scalar spectrum. This spectrum does not contain masses below the BF bound at any level in the KK tower of spin-$0$ fluctuations. Therefore, it presents an example of a non-supersymmetric, yet perturbatively stable, solution of ten-dimensional massive IIA supergravity. We recall that this solution is a well-defined background of massive IIA string theory \cite{Guarino:2015jca,Varela:2015uca}, and that the analysis of \cite{Guarino:2020jwv} has moreover excluded D$p$-brane-jet instabilities for $p=2,4,6,8$\,. We have also presented evidence that six other non-supersymmetric AdS$_4 \times S^6$ backgrounds of massive IIA supergravity with smaller residual symmetry are also perturbatively stable in ten dimensions.

Since these AdS$_4$ vacua have already passed a number of non-trivial tests regarding their non-perturbative stability, they seem to provide counterexamples to the belief that there are no stable non-supersymmetric AdS vacua of string theory. Therefore, our findings here challenge the AdS Swampland Conjecture \cite{Ooguri:2016pdq}. If the Swampland Conjecture is to hold, there must be an alternative decay channel for these non-supersymmetric AdS$_4$ vacua, which it would be imperative to unearth.\\

\noindent\textbf{Acknowledgements}:  
We would like to thank David Andriot and Davide Cassani for helpful discussions. AG is supported by the Spanish government grant PGC2018-096894-B-100 and by the Principado de Asturias through the grant FC-GRUPIN-IDI/2018/000174. EM is supported by the Deutsche Forschungsgemeinschaft (DFG, German Research Foundation) via the Emmy Noether program ``Exploring the landscape of string theory flux vacua using exceptional field theory'' (project number 426510644).

\section*{Appendix: $\mathbb{Z}_{2}^{3}$-invariant sector and the 56-bein}
\label{Appendix:6_vacua}

The $\mathbb{Z}_{2}^{3}$-invariant sector of the ISO(7) supergravity retains fields that are invariant under the action of three discrete $\mathbb{Z}^{(1,2,3)}_{2}$ transformations. When acting on an SL(8) vector $v_{\cI}$, $\cI=1, \dots, 8$, these induce reflections of the form $\mathbb{Z}^{(1)}_{2}: v_{4,5,6,7} \rightarrow -v_{4,5,6,7}$, $\,\mathbb{Z}^{(2)}_{2}: v_{2,3,6,7} \rightarrow -v_{2,3,6,7}$ and $\mathbb{Z}^{(3)}_{2}: v_{2,4,6,8} \rightarrow -v_{2,4,6,8}$. At the bosonic level, the set of invariant fields consists of the metric $\mathring{g}_{\mu\nu}$ and fourteen spin-$0$ real fields: $7$ scalars $\textrm{Im}z_{i}$ and $7$ pseudo-scalars $\textrm{Re}z_{i}$ with $i=1,\ldots,7$. The scalars are associated with Cartan generators of the E$_{7(7)}$ algebra given by
\begin{equation}
	\label{gen_scalars}
	\begin{array}{llll}
		g_{i}  = \displaystyle\sum_{\cI}  c_{i,\cI} \,t_{\cI}{}^{\cI} & ,
	\end{array}
\end{equation}
with
\begin{equation}
	c_{i,\cI} = \left(  
	\begin{array}{cccccccc}
		-1&-1&-1&1&1&1&1&-1 \\
		-1&1&1&-1&-1&1&1&-1 \\
		-1&1&1&1&1&-1&-1&-1 \\
		1&-1&1&1&-1&1&-1&-1 \\
		1&1&-1&-1&1&1&-1&-1 \\
		1&1&-1&1&-1&-1&1&-1 \\
		1&-1&1&-1&1&-1&1&-1
	\end{array}
	\right) \,,
\end{equation}
whereas the pseudo-scalars are associated with positive roots 
\begin{equation}
	\label{gen_pseudo-scalars}
	\begin{split}
		\tilde{g}_{1}  &=   t_{1238} \,, \,\,\, 
		\tilde{g}_{2}  =   t_{1458} \,, \,\,
		\tilde{g}_{3}  =   t_{1678}  \,,\\
		\tilde{g}_{4}  &=   t_{2578} \,, \,\,\,
		\tilde{g}_{5}  =   t_{4738}  \,, \,\,\,
		\tilde{g}_{6}  =   t_{6358}  \,, \,\,\,
		\tilde{g}_{7}  =   t_{8246} \,.
	\end{split}
\end{equation}
In \eqref{gen_scalars} and \eqref{gen_pseudo-scalars} we made use of the SL(8) branching of the adjoint of E$_{7(7)}$, namely,
\begin{equation}
	t_\alpha = \left\{ t_\cI{}^\cJ,\, t_{[\cI\cJ\cK\cL]} \right\}\,, \quad  \alpha = 1, \ldots, 133 \,,
\end{equation}
with $t_\cI{}^\cI = 0$. Then, the coset representative in \eqref{56-bein} is obtained upon direct exponentiation of \eqref{gen_scalars} and \eqref{gen_pseudo-scalars}
\begin{equation}
	\label{M_scalar_N=1}
	\mathcal{V}_{\underline{M}}{}^{\underline{A}}(z_{i})  =e^{12 \, \sum \textrm{Re}z_{i} \tilde{g}_{i}}
	\,\,\,
	e^{-\frac{1}{4} \, \sum \log(\textrm{Im}z_{i}) g_{i}} \ .
\end{equation}

Finally, the action of $\mathbb{Z}^{(1)}_{2} \times \mathbb{Z}^{(2)}_{2}$ on the eight gravitini of the maximal theory leaves only two of them invariant. Then modding out by the additional $\mathbb{Z}^{(3)}_{2}$ element leaves only one invariant gravitino. As a result, the $\mathbb{Z}_{2}^3$-invariant sector of the maximal ISO(7) supergravity describes the $\mathcal{N}=1$ supergravity coupled to seven chiral multiplets presented in the main text.

\providecommand{\href}[2]{#2}\begingroup\raggedright\endgroup

\end{document}